\documentclass[aps,twocolumn,showpacs]{revtex4}
\usepackage{graphicx}  
\begin{document}

\title{
Graphene Bilayer
Field-Effect Phototransistor for Terahertz and Infrared Detection
}
\author{V.~Ryzhii\footnote{Electronic mail: v-ryzhii(at)u-aizu.ac.jp} 
and M.~Ryzhii}
\address{
Computational Nanoelectronics Laboratory, University of Aizu, 
Aizu-Wakamatsu, 965-8580, Japan\\
Japan Science and Technology Agency, CREST, Tokyo 107-0075, Japan
}

\begin{abstract}
A graphene bilayer phototransistor (GBL-PT) is proposed and 
analyzed. The GBL-PT under consideration
has the structure of a field-effect transistor  with a GBL as
the  channel and the back and top gates. The positive bias of 
the back gate results in the formation of conducting
source and drain sections in the channel, while the negatively biased top
gate provides the potential barrier which is controlled by the charge
of the photogenerated holes. The features of the GBL-PT operation are
associated with the variations of both the potential distribution
and the energy gap in different sections of the channel
when the gate voltages and the charge in the barrier section change.
Using the developed GBL-PT device model, the spectral characteristics,
dark current,
responsivity
and detectivity are calculated as functions
of the applied voltages, energy of incident photons,
intensity of electron and hole scattering, 
and geometrical parameters. It is shown that the GBL-PT spectral
characteristics  are voltage tuned. The GBL-PT performance as photodetector
in the terahertz and infrared photodetectors can markedly exceed
the performance of other photodetectors.
\end{abstract}


\maketitle
\newpage
\section{Introduction}

At present,  infrared detectors are mostly based on 
narrow-gap semiconductors
 utilizing the interband transitions.
Technologies utilizing  HgCdTe and InSb are well developed for infrared 
detection and imaging~\cite{1,2}. 
The necessity of further extension of the wavelength range 
covered by photodetectors and imaging devices on their base, 
widening of their functionality,
as well as 
 cost reduction of the production by using a mature  processing 
technology has stimulated the development of quantum-well 
infrared photodetectors (QWIPs) based on A$_3$B$_5$ compound
systems and SiGe alloys and utilizing
intersubband
(intraband) transitions
(see, for instance,~\cite{2,3}). 
Quantum-dot and quantum-wire infrared photodetectors (QDIPs and QRIPs)
were also
proposed~\cite{4,5} and realized by many groups.
The utilization of  graphene layers and graphene bilayers~\cite{6,7}
opens up real prospects in the  creation of novel
 photodetectors.
The most important advantage of graphene   relates to 
the possibility to control in a wide range the energy gap 
by patterning of the graphene layer into an array of narrow strips
(nanoribbons)~\cite{8,9}. The energy gap in graphene bilayers can be varied
by the transverse electric field~\cite{10,11,12,13} 
in different gated heterostructures. 
The graphene-based 
photodetectors  can exhibit relatively high quantum efficiency
(due to the use of interband transitions) and 
 be easily
integrated with silicon readout circuits.
A photodetector for terahertz (THz) and infrared (IR) radiation
based on a field-effect transistor  structure with the channel
consisting of an array of graphene nanoribbons was proposed and analyzed
recently~\cite{14}. In this paper, we discuss the concept of
a THz/IR photodetector with the structure of a field-effect transistor 
with a graphene bilayer as
the device channel and photosensitive element. 
Using the developed device model, we
 calculate and analyze the detector characteristics. 
We demonstrate that such a graphene bilayer phototransistor (GBL-PT) 
can operate as very sensitive
and voltage tunable  THz/IR photodetector at elevated temperatures.


\section{Model}

The GBL-PT under consideration
has a structure similar to that of a GBL-field-effect transistor~\cite{15}
 shown schematically in Fig.~1a.
The GBL channel placed over a highly conducting substrate
is supplied with the 
source and drain contacts. The substrate plays the role of the back gate
which provides the formation of a two-dimensional electron gas (2DEG) in the channel
when the back gate is biased positively with respect to the source and drain:
$V_b > V_d > 0$.
There is a top electrode serving as the top gate 
which is biased negatively ($V_t < 0$). 
Here, $V_b$, $V_t$, and $V_d$ are the back-gate,  top-gate, 
and source-drain
voltages, respectively.
The negative bias of the top gate results in
a depletion  of the  section of the channel beneath the top
gate (which in the following
are referred to as the gated section), so that
the channel is partitioned into two  highly conducting 
sections (source and drain sections)
and the  depleted gated section.
 In the gated section
the potential barrier for electrons is formed. 
This barrier controls the injected 
electron
current from the source to drain. The GBL-PT band diagram 
 under
the bias voltages corresponding to the operation conditions is shown
in Fig.~1b. We shall 
assume that the back gate (substrate) 
and the top gate are sufficiently transparent for the incoming radiation.
The GBL-PT operation  is associated with the variation of the source-drain
electron current under illumination when the electron-holes pairs
are generated in the depleted sections. The photogenerated 
electrons are swept out to the conducting section 
(as shown schematically in Fig.~1(b)), whereas the 
photogenerated holes accumulated in the deleted section result in 
lowering of the potential
barrier for the injected electrons.
As shown, the variation of the injected electron current can substantially 
exceed
the current created by entirely photogenerated electrons and holes, so that
GBL-PTs can exhibit large photoelectric gain.
\begin{figure}[t]
\vspace*{-0.4cm}
\begin{center}
\includegraphics[width=8.0cm]{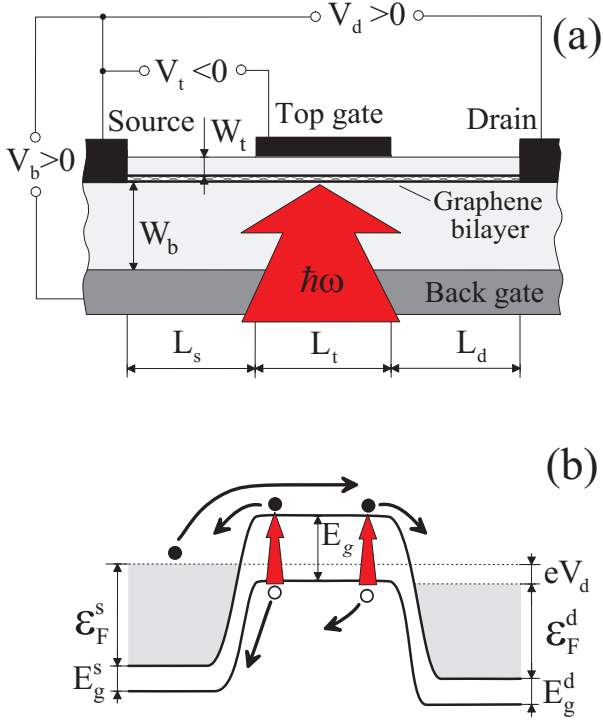}
\caption{Schematic view of (a) a GBL-PT structure and (b) its band diagrams
at bias voltages. Dark arrows indicate directions
of electron (opaque circles) and holes (open circles) propagation.
}
\end{center}
\end{figure}
We consider the situation when the gated section is fully depleted.
It occurs when $ V_{th}^* < V_t < V_{th}$,
where $e = |e|$ is the electron charge and
$V_{th}^* < V_{th} < 0$ are the threshold voltages.
The latter correspond to the top-gate voltages at which the bottom
of the valence band and the
conduction band in the gated section align with the  Fermi
level in the source section ($V_{th}^* = V_{th} - 2E_g/2$).
The energy gaps in the source, gated, and drain regions,
$E_g^s$, $E_g$, and $E_g^D$, as well as 
  threshold voltages are given by

\begin{equation}\label{eq1}
E_g^s = \frac{ eV_bd}{2W},
\,
E_g = \frac{ e(V_b - V_t)d}{2W}, 
\,
E_g^d = \frac{ e(V_b - V_d)d}{2W},
\end{equation}
\begin{equation}\label{eq2}
V_{th} \simeq - V_b\biggl(1 + \frac{a_B}{4W}\biggr),\,
V_{th}^* \simeq - V_b\biggl(1 + \frac{a_B}{4W} + \frac{2d}{W}\biggr),
\end{equation}
where $e = |e|$ is the electron charge,$d$ is the spacing between the graphene layers in the GBL,
$W$ is the thickness of the layers separating the GBL and the gates
(which are assumed to be equal to each other: $W_t = W_b = W$),
$a_B = k\hbar^2/me^2$ is the Bohr radius, $k$ is the dielectric constant,
$m$ is the effective mass of electrons and holes in the GBL,
and $\hbar$ is the Planck constant. The parameters $d_0/W$ and $a_B/4W$
are assumed to be small. For simplicity, we shall neglect some difference
between $d$ and the efficient spacing $d_s$ which accounts for
the effects of electron screening of the transverse electric field
between the graphene layer~\cite{10,11}.

Considering that electrons in the source and drain sections form
 2DEGs  (generally with different densities),
the Fermi energies in these sections can be presented as
%
%
%
$$
\varepsilon_F^s \simeq k_BT 
\ln\biggl[
\exp\biggl(\frac{a_B}{8W}\frac{eV_b}{k_BT}\biggl) - 1\biggr],
$$
\begin{equation}\label{eq3}
\varepsilon_F^d = k_BT 
\ln\biggl[
\exp\biggl(\frac{a_B}{8W}\frac{e(V_b - V_d)}{k_BT}\biggl) - 1\biggr], 
\end{equation}
where $k_B$ is the Boltzmann constant and $T$ is the temperature.
In particular,  at sufficiently high back-gate voltages, namely at 
$eV_b > (8W/a_B)k_BT$, one obtains
$\varepsilon_F^s 
\simeq eV_b(a_B/8W)$ and $\varepsilon_F^d 
\simeq e(V_b - V_d)(a_B/8W)$. 
One can see that $E_g^{s,d}/\varepsilon_F^{s,d} \simeq 4d/a_B$.
The latter value is rather small in the case of gate layers made of SiO$_2$
and is very small in the case of gate layers with elevated
dielectric constant (say, HfO$_2$).


\section{Dark current}

The source-drain current created by the electrons injected from 
the source and
drain section into the gated section is given by
\begin{equation}\label{eq4}
J = \beta\,J_m\biggl[\exp\biggl(\frac{\varepsilon_F^s - \Delta^s}{k_BT}\biggr)
- \exp\biggl(-\frac{\varepsilon_F^d - \Delta^d}{k_BT}\biggr)\biggr].
 \end{equation} 
Here  $\Delta^s$ and  $\Delta^d$ are  the heights of the potential
barriers for electrons from the source and drain sides, respectively,
and
\begin{equation}\label{eq5}
J_m = e\frac{\sqrt{2m}(k_BT)^{3/2}}{\pi^{3/2}\hbar^2}.
\end{equation}
The factor $\beta$ is the fraction of the injected electrons passed
through the gated section despite their scattering on impurities
and acoustic phonons. Solving the 2D kinetic Boltzmann equation
for the electron distribution function in this section, one can obtain
 $\beta \simeq 1$
in the ballistic regime of the electron transport across the gated section
($\nu\tau < 1$ or $L_t \ll \sqrt{2k_BT/m}/\nu$),
and $\beta \simeq \sqrt{\pi}/\nu\tau$
in the collision-dominated regime ($\nu\tau \gg 1$),  where
$\tau = L_t\sqrt{m/2k_BT}$ is the effective
 ballistic transit time across the gated section of electrons
with the thermal velocity $v_T = \sqrt{2k_BT/m}$, and $L_t$ is the length
of the top gate~\cite{15}.

In the dark conditions, $\Delta^s = \Delta_0^s$ and  $\Delta^d = \Delta_0^d$
with
$$
 \Delta_0^s = -\frac{e(V_b + V_t)}{2},\qquad 
\Delta_0^d = -\frac{e(V_b + V_t)}{2} - \varepsilon_F^s +
\varepsilon_F^d + eV_d 
$$
\begin{equation}\label{eq6}
\simeq -\frac{e(V_b + V_t)}{2}
+ \biggl[1 - \biggl(\frac{a_B}{8W}\biggr)\biggr]eV_d.
 \end{equation} 
Using Eqs.~(3), (4), and (6), we arrive at the following formula for the
dark current:
$$
J_0 \simeq \beta\,J_m\biggl[\exp\biggl(\frac{a_B}{8W}\frac{eV_b}{k_BT}\biggr) - 1\biggr]
$$
$$
\exp\biggl[\frac{e(V_b + V_t)}{2k_BT}\biggr]
\biggl[1
- \exp\biggl(-\frac{eV_d}{k_BT}\biggr)\biggr]
$$
\begin{equation}\label{eq7}
\simeq  \beta\,J_m
\exp\biggl[- \frac{e(V_{th}  -  V_t)}{2k_BT}\biggr]\biggl[1
- \exp\biggl(-\frac{eV_d}{k_BT}\biggr)\biggr].
 \end{equation} 
%
\begin{figure}[t]
\vspace*{-0.4cm}
\begin{center}
\includegraphics[width=8.0cm]{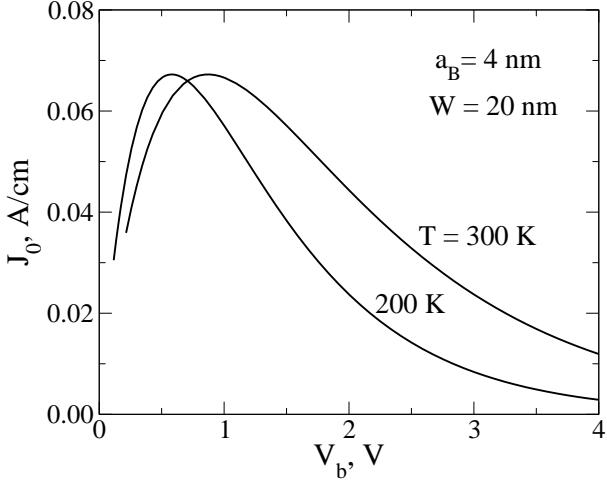}
\caption{
Dark carrent $J_0$ as a function of back-gate voltage $V_b$ 
(at $V_t = -V_b[1 + (a_B/4 + 2d)/W$)  
 at different temperatures~$T$.
}
\end{center}
\end{figure}
Figure~2 shows the dependence of the dark current $J_0$ on the
back-gate voltage $V_b$ provided that the top-gate voltage is maintained
to be
$V_t = V_{th}^* = -V_b[1 + (a_B/4 + 2d)/W]$.  
This corresponds to the highest barrier in 
the gated section at which  the interband tunneling can still be neglected, in particular,
owing to a small density of thermal holes in this section.
It is assumed that  $d = 0.36$~nm, $a_B = 4$~nm,  $W = 20$~nm,  
$\beta = 0.1$,
and $V_d > k_BT/e$.
Nonmonotonic behavior of the dark current-voltage characteristics
shown in Fig.~2 can be attributed to the interplay of an increase
of electron density in the source (drain) section 
with
increasing $V_b$ 
and an increase  in the height of the potential barrier
in the gated section when both $V_b$ and $|V_t|$ simultaneously increase.  



\section{Photocurrent and responsivity}

As a result of illumination with the photon energy $\hbar\omega > E_g$,
the photogenerated holes are accumulated in the gated section.
Their density $\Sigma$ can be found from the following equation
governing the balance between the photogeneration of holes and their
escape to the source and drain sections:
%
%
$$
G_{\omega} = \frac{\beta_cJ_m}{e} \frac{\Sigma}{\Sigma_t}
\exp\biggl(- \frac{d}{2W}\frac{eV_t}{k_BT}\biggr)
$$
\begin{equation}\label{eq8}
\exp\biggl[\frac{e(V_b + V_t)}{2k_BT}\biggr]
\biggl[1 + \exp\biggl(-\frac{eV_d}{k_BT}\biggr)\biggr].
\end{equation} 
Here $\beta_c$ is the fraction of the holes
injected from the gated section into the source and drain sections
(i.e., into the contact sections)
but not reflected back owing to the scattering 
($\beta_c$ is determined not only
by the hole collision frequency  in these section but also by
the rate of recombination in these sections and the contacts),
$\Sigma_t = 2mk_BT/\pi\,\hbar^2$,
and
$G_{\omega}$ is the rate of photogeneration
of electrons and holes owing to the absorption of the incident THz/IR
radiation. This quantity depends on the intensity of
 radiation $I_{\omega}$, the absorption quantum efficiencies
$\alpha_{\omega}$, $\alpha^s_{\omega}$, and $\alpha^s_{\omega}$
in the pertinent sections, and their lengths.
The  absorption quantum efficiencies in question
are  given by~\cite{16}
\begin{equation}\label{eq9}
\alpha_{\omega} =
\biggl(\frac{\pi\,e^2}{c\hbar}\biggr)
\biggl(\frac{\hbar\omega + 2\gamma_1}{\hbar\omega + \gamma_1}\biggr)
\Theta(\hbar\omega - E_g),
\end{equation} 
\begin{equation}\label{eq10}
\alpha^{s,d}_{\omega} \simeq
\biggl(\frac{\pi\,e^2}{c\hbar}\biggr)
\biggl(\frac{\hbar\omega + 2\gamma_1}{\hbar\omega + \gamma_1}\biggr)
\Theta(\hbar\omega - E_g^{s,d})\,B_{\omega},
\end{equation} 
where $c$ is the speed of light,  $\gamma_1 \simeq 0.4$~eV
is the band parameter,~\cite{12} and $\Theta(\hbar\omega)$
is the unity step function reflecting the energy dependence of the
density of states near the fundamental edge of absorption.
To take into account some smearing $\gamma$ of this edge, we set 
$\Theta(\hbar\omega) = [1/2 + (1/\pi)\tan^{-1}(\hbar\omega/\gamma)]$.   
The factor $B_{\omega}$ in Eq.~(10) reflects 
the Burstein-Moss effect~\cite{17}:
$$
B_{\omega} = \biggl[1 + 
\exp\biggl(- \frac{\hbar\omega - E_g^{s,d} - 2\varepsilon_F^{s,d}}{2k_BT}
\biggr)\biggr]^{-1} 
$$
$$
\times\biggl[1 + 
\exp\biggl(- \frac{\hbar\omega +E_g^{s,d} + 2\varepsilon_F^{s,d}}{2k_BT}
\biggr)\biggr]^{-1}
$$
$$
\simeq \biggl[1 + 
\exp\biggl(- \frac{\hbar\omega - E_g^{s,d} - 2\varepsilon_F^{s,d}}{2k_BT}
\biggr)\biggr]^{-1}.
$$
For  the THz/FIR radiation with   $E_g \lesssim \hbar\omega \ll \gamma_1$,
Eq.~(9) yields
$\alpha_{\omega} \simeq \alpha = 2\pi\,e^2/c\hbar = 2\pi/137$.
However, the absorption coefficient in the source and drain section
can be rather small due to the Burstein-Moss effect.
This occurs if the 2D electron gas in these sections is degenerate
($\varepsilon_F^{s,d} \gg k_BT$) and the photon energy does not markedly
exceeds the energy gap..
Indeed, at $ \hbar\omega \gtrsim  E_g \simeq 2 E_g^{s,d}$, 
considering that $E_g^{s,d}/\varepsilon_F^{s,d} \simeq 4d/a_B  <  1$,
from Eq.~(10) one obtains
$\alpha^{s,d}_{\omega} \simeq  
\alpha\exp(-\varepsilon_F^{s,d}/k_BT) 
= \alpha\exp[- (a_B/8W)(eV_b/k_BT)] \ll \alpha$.
Disregarding therefore the absorption of radiation
in the source and drain section and considering that
\begin{equation}\label{eq11}
G_{\omega} = \frac{L_t\alpha_{\omega}I_{\omega}}{\hbar\omega},  
\end{equation} 
Eq.~(9) can be presented as
%
$$
L_t\frac{\alpha_{\omega}\,I_{\omega}}{\hbar\omega} = \frac{\beta_cJ_m}{e}
\frac{\Sigma}{\Sigma_t}\exp\biggl(-\frac{d}{2W}\frac{eV_t}{k_BT}\biggr)
$$
\begin{equation}\label{eq12}
\exp\biggl[\frac{e(V_b + V_t)}{2k_BT}\biggr]
\biggl[1 + \exp\biggl(-\frac{eV_d}{k_BT}\biggr)\biggr].
\end{equation} 
\begin{figure}[t]
\vspace*{-0.4cm}
\begin{center}
\includegraphics[width=8.0cm]{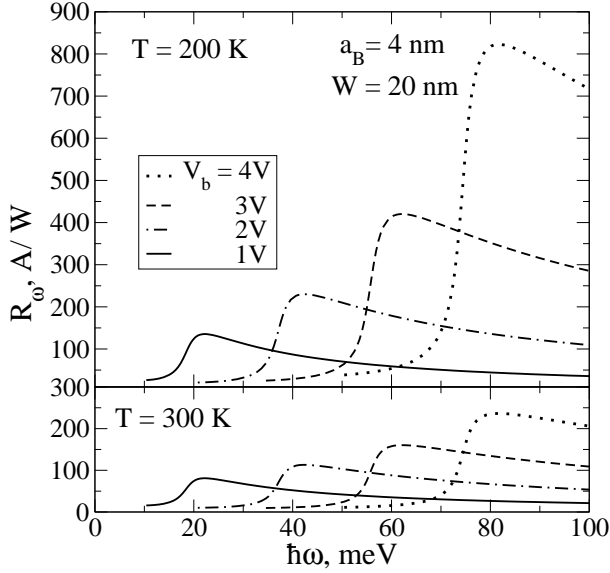}
\caption{
GBL-PT responsivity $R_{\omega}$ versus photon energy 
$\hbar\omega$
 for different
back-gate voltages $V_b$ at different temperatures $T$.
}
\end{center}
\end{figure}
\begin{figure}[t]
\vspace*{-0.4cm}
\begin{center}
\includegraphics[width=8.0cm]{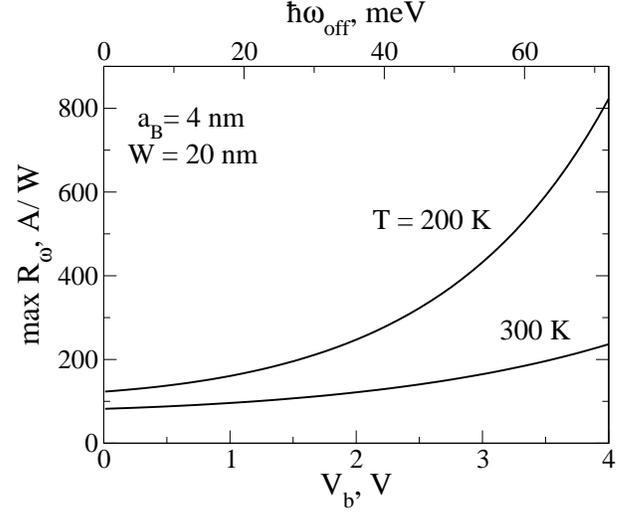}
\caption{
Voltage dependences of maximum GBL-PT responsivity 
at different temperatures $T$.
}
\end{center}
\end{figure}
The charge of the photogenerated holes in the gated section
gives rise to lowering of the potential barrier
by the value
\begin{equation}\label{eq13}
\Delta^{s,d} - \Delta_0^{s,d} = -\frac{4\pi\,e^2W}{k}\Sigma.
\end{equation}
Considering Eqs.~(4), (7), and  (13), the variation of the source-drain
current under illumination $\Delta\,J =  J - J_0$, i.e., the value of the photocurrent
can be presented by   the following
formula:
\begin{equation}\label{eq14}
\Delta J = J_0\frac{4\pi\,e^2W}{kk_BT}\Sigma.
\end{equation}
It should be noted that the contribution of the photogenerated
electrons to the net photocurrent can be neglected because is is small
and the photogenerated electrons are swept out from the gated section
to both the source and drain section (virtually in equal portions).
Then, using Eqs.~(7), (11), and (14), we arrive at

$$
\Delta J =  e\frac{L_t\beta}{\beta_c}
\biggl( \frac{8W}{a_B}\biggr)
\biggl[\exp\biggl(\frac{a_B}{8W}\frac{eV_b}{k_BT}\biggr) - 1\biggr]
$$
\begin{equation}\label{eq15}
\exp\biggl(\frac{d}{2W}\frac{eV_t}{k_BT}\biggr)
\biggl[\frac{1 - \exp(-eV_d/k_BT)}{1 + \exp(-eV_d/k_BT)}\biggr]
\frac{\alpha_{\omega}\,I_{\omega}}{\hbar\omega}.
\end{equation}

Using Eq.~(15),
the GBL-PT responsivity defined as $R_{\omega} = \Delta J/(L_t + 2L_c)I_{\omega} $,
where $L_t + 2L_c$ is the net length of the GBL channel 
(the lengths of the source and drain sections
are assumed to be equal to each other:$L_s = L_d = L_c$),
can be presented
$$
R_{\omega} =  \frac{e\alpha\,C}{\hbar\omega}\,
\Theta\biggl(\hbar\omega - \frac{ed(V_b - V_t)}{2W}\biggr)
\displaystyle\biggl[\exp\biggl(\frac{a_B}{8W}\frac{eV_b}{k_BT}\biggr) - 1\biggr]
$$
\begin{equation}\label{eq16}
\times \biggl(\frac{8W}{a_B}\biggr)\exp\biggl(\frac{d}{2W}
\frac{eV_t}{k_BT}\biggr)
\biggl[\frac{1 - \exp(-eV_d/k_BT)}{1 + \exp(-eV_d/k_BT)}\biggr].
\end{equation}
Here, 
$C = L_t\beta/(L_t + 2L_c)\beta_c$ is the collision factor.
At $\hbar\omega = 10$~meV and $C = 1$, one obtains
$\overline{R_{\omega}} = (e\alpha\,C/\hbar\omega) \simeq 4.6$~A/W.

In the most interesting situation when the 2D gases 
in the source and drain
sections are degenerate, the top gate voltage is chosen to 
provide relatively high barrier for electrons in the source 
and drain sections,
and the source-drain voltage is sufficiently large, i.e., at 
$eV_b > k_BT(8W/a_B)$,  $V_t = V_{th}^* = - V_b[1 + (a_B/4 + 2d)/W]$, and 
$eV_d > k_BT$, Eq.~(16) can be reduced to the following:
$$
R_{\omega} \simeq  \frac{e\alpha\,C}{\hbar\omega}\,
\Theta(\hbar\omega - \hbar\omega_{off})
\biggl(\frac{8W}{a_B}\biggr)
$$
\begin{equation}\label{eq17}
\times\biggl[\exp\biggl(\frac{a_B}{8W}\frac{eV_b}{k_BT}\biggr)
\biggl(1 - \frac{4d}{a_B}\biggr)\biggr].
\end{equation}
Here $\hbar\omega_{off} = E_g
\simeq ed(V_b - V_t)/2W \gtrsim edV_b/W$ is the photon  
cut-off energy
at which
the GBL-PT responsivity reaches a maximum: 
\begin{equation}\label{eq18}
{\rm max} R_{\omega} \simeq  \frac{\alpha\,C}{V_b}
\biggl(\frac{8W^2}{da_B}\biggr)
\biggl[\exp\biggl(\frac{a_B}{8W}\frac{eV_b}{k_BT}\biggr)
\biggl(1 - \frac{4d}{a_B}\biggr)\biggr].
\end{equation}

Figure~3 shows the GBL-PT responsivity $R_{\omega}$
as a function of the photon energy $\hbar\omega$ 
calculated using Eq.~(16) for different
back-gate voltages $V_b$ at different temperatures.
Here and in the following figures,
it is set that for each value of the back-gate voltage $V_b$
the top gate voltage  is chosen to be $V_t = V_{th}^*
= - V_b[1 + (a_B/4 + 2d)/W]$. 
The source-drain voltage is assumed to be $V_d > k_BT/e$.
In such a case, the thermionic dark current
is lowered while the interband tunneling is still negligible.
We assume that  $d = 0.36$~nm,
$a_B = 4$~nm,   $W = 20$~nm,  $\gamma = 2$~meV, $\beta = 0.1$, 
and $C = 1$. The spectral dependences shown in Fig.~3 correspond to
the cut-off photon energies $\hbar\omega = E_g \simeq edV_b/W$
(when  $|V_t| \gtrsim V_b$).

Figure~4 shows the dependences of the responsivity maximum value
max~$R_{\omega}$ on the back-gate voltage $V_b $ (and the photon energy
$\hbar\omega_{off}$)
for the same parameters 
as in Fig.~3.

\section{Photoelectric gain and detectivity}

Considering  Eq.~(15) and taking into account that 
the photocurrent created
by the photogenerated electrons and holes as such is equal to 
$\Delta J_0 =  eL_t\alpha_{\omega}I_{\omega}/\hbar\omega$, 
the photoelectric gain $g = \Delta J/\Delta J_0$ can be presented as 
(compare with Eq.~(18))
\begin{equation}\label{eq20}
g \simeq  \frac{\beta}{\beta_c}
\biggl( \frac{8W}{a_B}\biggr)
\exp\biggl[\biggl(\frac{a_B}{8W}\frac{eV_b}{k_BT}\biggr) 
\biggl(1 - \frac{4d}{a_B}\biggr)\biggr],
\end{equation}
where all the factors in the right-hand side exceed or greatly exceed
unity 
if $4d/a_B < 1$ ($4d/a_B \simeq 0.36$ and 0.07 in the case of 
SiO$_2$ and HfO$_2$ gate layers, respectively.)

Calculating 
the GBL-PT dark current limited detectivity 
as $D^* = R_{\omega}/\sqrt{4egJ_0/H}$,
where $H$ is the GBL-PT width 
(in the direction perpendicular to the current), at properly chosen
relationship between $V_b$ and $V_t$ (as above),
we arrive at the following
formula:
\begin{equation}\label{eq21}
D^* \simeq  \frac{e\alpha\,C^*}{\hbar\omega_{off}}
\sqrt{\frac{H}{4eJ_m}}
\sqrt{\frac{8W}{a_B}}
\exp\biggl(\frac{\hbar\omega_{off}}{nk_BT}\biggr), 
\end{equation}
where $ C^* = C \sqrt{\beta_c}/\beta =[L_t/(L_t + 2L_c)\sqrt{\beta_c}]$
and $n = (16d/a_B)/[1 + 4d/a_B]$.
A point worth noting is that the factor $n$ in the exponential
dependence in Eq.~(21) can be  about or smaller than unity. 
Indeed, for $a_B = 4 - 20$~nm, one obtains $n \simeq 0.27 - 1 $.
This provides fairly steep increase in $D^*$ with 
increasing $\hbar\omega_{off}$.
\begin{figure}[t]
\vspace*{-0.4cm}
\begin{center}
\includegraphics[width=8.0cm]{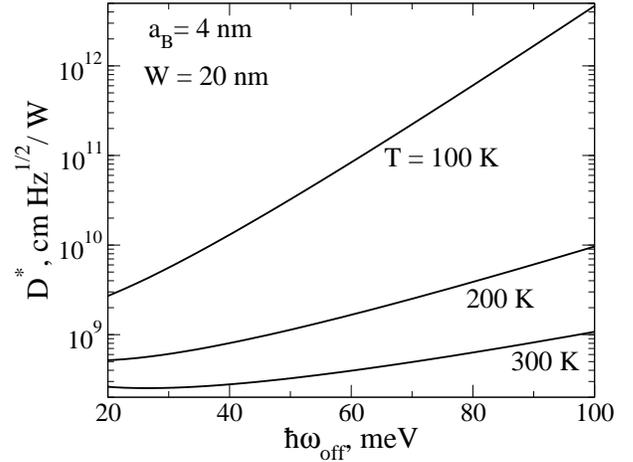}
\caption{GBL-PT dark current limited detectivity   $D^*$
versus cut-off photon energy $\hbar\omega_{off}$ at optimized
gate voltages, $V_b$ and $V_t$, and   different different temperatures
$T$.
}
\end{center}
\end{figure}
%
\begin{figure}[t]
\vspace*{-0.4cm}
\begin{center}
\includegraphics[width=8.0cm]{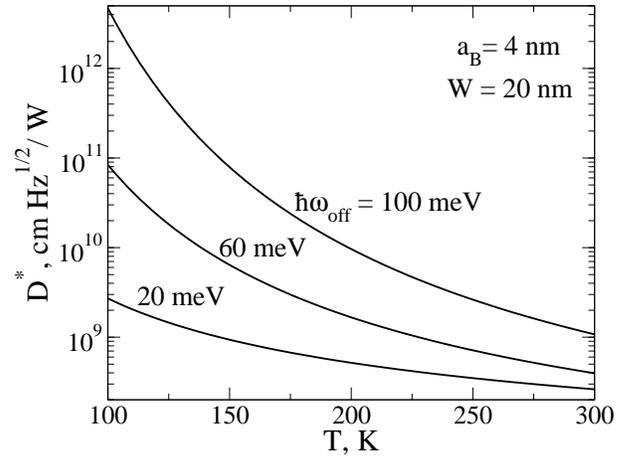}
\caption{
Temperature dependences of GBL-PT dark current 
limited detectivity $D^*$ for different  
 cut-off photon energies $\hbar\omega_{off}$.
} 
\end{center}
\end{figure}

\begin{figure}[t]
\vspace*{-0.4cm}
\begin{center}
\includegraphics[width=8.0cm]{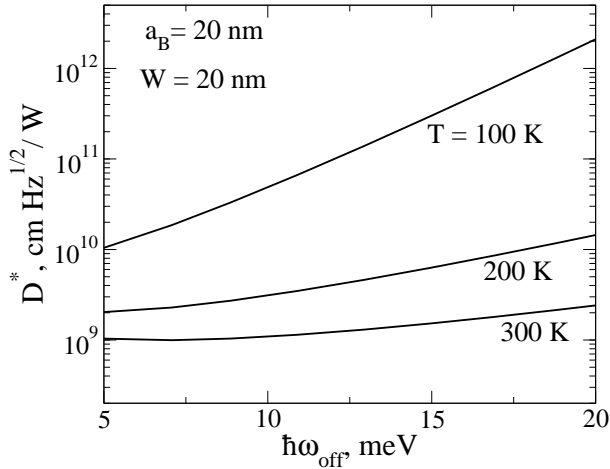}
\caption{The same as in Fig.~5 but for GBL-PT with different
value of the Bohr radius $a_B$ (different dileltric constant 
of gate layers $k$). 
}
\end{center}
\end{figure}
Figures~5 and 6 demonstrate   the
dark current limited detectivity $D^*$ (under the optimized conditions)
as a function of the cut-off photon energy $\hbar\omega_{of}$ at different temperatures $T$
and as a function of $T$  at given values $\hbar\omega_{off}$.
One can see that  $D^*$ can be fairly large even at room temperatures.
The detectivity markedly decreases with decreasing cut-off photon energy.
However, as shown in Fig.~7,
in GBL-PTs with relatively high-$k$ gate layers in which
the Bohr radius can be large, a rather high detectivity can be achieved
in the range of low cut-off photon energies, in particular, 
those corresponding
to the THz range of spectrum.


\section{Comments}

As follows from Eq.~(16) - (20) and demonstrated in Figs.~3 - 7,
the GBL-PT responsivity and detectivity
can be very large even at room temperatures
exceeding those of QWIP and QDIPs. 
This is attributed to the following.
First of all, the quantum efficiency of the interband
$\alpha$ is relatively large
(in comparison, say, with the intersubband transitions
in single QWIPs).
Second, the photoelectric gain exhibited by GBL-PTs can also be very
large. This is associated with a higher energy barrier 
(high activation energy)
 for the photogenerated
holes accumulated in the gated section  in comparison with
the activation energies for electrons in the source and drain sections.
The difference between this  activation energies is equal to 
$\varepsilon_F^{s,d} - E_g + E_g^{g,s} \simeq (a_B/8W)(1 - 4d/a_B)eV_b $.
An increase in the gate voltages results in an increase of the 
Fermi energy of electrons in the source and drain section and, 
in a rise of the
activation energy
 for the photogenerated
holes leading to an increase of their lifetime.
As a result, the temperature dependence of the GBL-PT detectivity
is given by the factor $\exp(\hbar\omega_{off}/nk_BT)$ with $n < 2$
in contrast to QWIPs (optimized)
for which
$D^* \propto  \exp(\hbar\omega_{off}/2k_BT)$  
(see, for instance, Refs.~\cite{3,5}). This might open prospects to
use GBL-PTs at elevated temperatures.

The GBL-PT responsivity and detectivity can be limited by the interband
tunneling of the photogenerated holes if the width of the junction
between the source (or drain) section and the gated section $W^*$
is too small ($W^*$ depends on $W$ as well as $V_b$ and $V_t$).
This can deteriorate the GBL-PT performance
due to a decrease in the photoelectric
gain associated with a shortening of the lifetime of 
the photogenerated holes.
Estimating  the probability of interband tunneling as~\cite{18}
$\exp(- \pi\,m^{1/2}E_g^{3/2}/
2\sqrt{2}e\hbar{\cal E})$, where ${\cal E} \simeq E_g/eW^*$,
we arrive at the following condition when the tunneling 
of the photogenerated holes might be essential: 
\begin{equation}\label{eq21}
W^* \gtrsim  \frac{a_B}{d}\frac{\hbar}{\pi\,\sqrt{2m}}\frac{\sqrt{\hbar\omega_{off}}}{k_BT}.
\end{equation}
For instance, for $a_B = 4$~nm, $\hbar\omega_{off} = 10 - 100$~meV,
and $T = 300$~K, inequality (21) yields $W^* \gtrsim 12  - 40$~nm.


The collision factor $C$ in Eqs.~(16) - (18) can influence
the GBL-PT performance. It
depends on the device geometrical parameters $L_t$
and $L_c$.
Since the propagation of holes in the source and drain sections
can be strongly affected by  scattering on electrons
due to their large density,  the hole collision frequency 
in the source and drain sections
$\nu_c > \nu$ (or even $\nu_c \gg \nu$).
In this case, $C$ can exceed unity.
This is because strong  collisions of holes in the source and drain section
can markedly decrease the current of the photogenerated holes from
the gated section into the source section (as well as into the drain section) increasing the holes lifetime and, hence, the photoelectric gain. 
To follow the dependence of $C$ (and, consequently, $R_{\omega}$) on
$L_t$ and $L_c$, 
 assuming that the holes recombine primarily
at the contacts,
one can use the following interpolation formulas:
$\beta_c = [1 + (\nu_c\,L_c/v_T)^2/\pi]^{-1/2}$.
and
$\beta = [1 + (\nu\,L_t/v_T)^2/\pi]^{-1/2}$.
As a result, we obtain
\begin{equation}\label{eq22}
C \simeq \frac{L_t}{(L_t + 2L_c)}\sqrt{\frac{1 + (\nu_c\,L_c/v_T)^2/\pi}{1 + (\nu\,L_t/v_T)^2/\pi}}.
\end{equation}
As follows from Eq.~(21),  $C$ as a function of $L_t$  
exhibits a maximum at a certain value of the latter.
If both $L_c$ and $L_t$ are large, so that
the electron transport in all the sections is collision dominated,
$C \simeq [L_c/(L_t + 2L_c)](\nu_c/\nu)$.

When the photon energy 
$\hbar\omega \gtrsim E_g^{s,d} + \varepsilon_F^{s,d}$,
the radiation absorption in the source and drain sections 
can be essential.
In such a  spectral range, the holes photogenerated in these section
can substantially affect the net hole charge in the gated section.
In this case,  the quantity $G_{\omega}$ in the right-hand side
of Eq.~(8) should be modified to take into account the extra 
holes photogenerated
in the source and drain sections and injected into the gated section.
The contribution of the holes photogenerated in the source and drain
section can result in a modification in the spectral characteristic
of the responsivity  at elevated  photon energies
according to the frequency dependence of the factor $B_{\omega}$
in Eq.~(10).
As a result,
a duplicated maxima of $R_{\omega}$ can appear which correspond
to $\hbar\omega \simeq \hbar\omega_{off}
\gtrsim edV_b/W$ and
 $\hbar\omega \simeq \hbar\omega_{off}/2 + (a_B/8W)eV_b
\gtrsim (edV_b/W)(1 + a_B/4d)/2 > \hbar\omega_{off}$.


We considered a GBL-PT with the structure of a single FET.
Actually, analogous GBL photodetectors can be made of multiple
periodic GBL-PT structures. Such photodetectors can surpass 
the GBL-PT considered above. However, their operation
can be complicated by additional features of the photogenerated
holes transport. As a result, the potential distribution along
the GBL channel can be nontrivial as 
it takes place in multiple QWIP (see, for instance, Refs~\cite{19,20}),
so that special studies of multiple GBL-PTs are required.

\section{Conclusions}

We proposed a GBL-PT and calculate its spectral characteristics, 
dark current, responsivity, and dark current limited detectivity.
It was shown that GBL-PTs with optimized structure at properly chosen
applied voltages can surpass the photodetectors of other types.
The main advantages of GBL-PTs are associated with the utilization of
interband transitions with relatively high quantum efficiency, 
high photoelectric gain, and possibility
of operation at elevated temperatures.
The advantages of the GBL-PT under consideration in comparison with
QWIPs, QDIPs, and QRIPs, 
 as well as with HgCdTe and InSb detectors can also
be easy fabrication and integration with silicon
(or graphene) readout circuits and the voltage tuning
of the spectral characteristics.

\section*{Acknowledgments}
The authors are grateful to S.~Brazovskii, N.~Kirova,
V.~Mitin, V.~Aleshkin, T.~Otsuji, and E.~Sano for useful discussions.
The work was supported by the Japan Science and 
Technology Agency, CREST,  Japan.

\end{document}